\documentclass[conference]{IEEEtran}
\IEEEoverridecommandlockouts

\usepackage{cite}
\usepackage{amsmath,amssymb,amsfonts}
\usepackage{algorithmic}
\usepackage{graphicx}
\usepackage{textcomp}
\usepackage{xcolor}
\usepackage{array}
\usepackage{multirow}
\usepackage{hyperref}
\usepackage[margin=1in]{geometry}
\def\BibTeX{{\rm B\kern-.05em{\sc i\kern-.025em b}\kern-.08em
    T\kern-.1667em\lower.7ex\hbox{E}\kern-.125emX}}
\begin{document}

\title{Extreme Learning Machine Based System for DDoS Attacks Detections on IoMT Devices\\
}
\author{
\IEEEauthorblockN{Nelly Elsayed}
\IEEEauthorblockA{\textit{School of Information Technology} \\
\textit{University of Cincinnati}\\
Cincinnati, Ohio \\
elsayeny@ucmail.uc.edu}
\and
\IEEEauthorblockN{Lily Dzamesi}
\IEEEauthorblockA{\textit{School of Information Technology} \\
\textit{University of Cincinnati}\\
Cincinnati, Ohio \\
dzamesly@mail.uc.edu}
\\  
\IEEEauthorblockN{Murat Ozer}
\IEEEauthorblockA{\textit{School of Information Technology} \\
\textit{University of Cincinnati}\\
Cincinnati, Ohio \\
ozermm@ucmail.uc.edu}
\and
\IEEEauthorblockN{Zag ElSayed}
\IEEEauthorblockA{\textit{School of Information Technology} \\
\textit{University of Cincinnati}\\
Cincinnati, Ohio \\
elsayezs@ucmail.uc.edu}
}

\maketitle

\begin{abstract}
The Internet of Medical Things (IoMT) represents a paradigm shift in the healthcare sector, enabling the interconnection of medical devices, sensors, and systems to enhance patient monitoring, diagnosis, and management. The rapid evolution of IoMT presents significant benefits to the healthcare domains. However, there is a rapid increase in distributed denial of service (DDoS) attacks on the IoMT networks due to several vulnerabilities in the IoMT-connected devices, which negatively impact patients' health and can even lead to deaths. Thus, in this paper, we aim to save lives via investigating an extreme learning machine for detecting DDoS attacks on IoMT devices. The proposed approach achieves a high accuracy at a low implementation budget. Thus, it can reduce the implementation cost of the DDoS detection system, making the model capable of executing on the fog level. 
\end{abstract}


\section{Introduction}

\begin{figure}[t]
\centerline{\includegraphics[width=8cm,height=5cm]{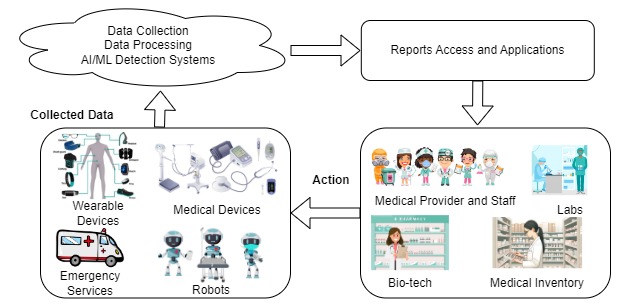}}
\caption{The Internet of Medical Things (IoMT) components.}
\label{IoMT_network}
\end{figure}

The healthcare industry is undergoing a significant transformation driven by the integration of digital technologies. A key component of this evolution is the Internet of Medical Things (IoMT), an extension of the broader Internet of Things (IoT) concept explicitly tailored for the medical domain~\cite{rani2023federated}. 
The Internet of Medical Things (IoMT) refers to the interconnected network of medical devices and applications that communicate over the internet \cite{joyia2017internet}. The IoMT connects a wide range of medical devices, sensors, software applications, and healthcare systems through interconnected network technologies, including wearable sensors that track vital signs, implants that regulate chronic conditions, and advanced hospital diagnostic tools \cite{dwivedi2022potential}. The IoMT components are shown in Figure~\ref{IoMT_network}. The primary function of these systems is to collect, transmit, and process health data to improve patient care  \cite{hameed2021systematic}. The IoMT enables the transfer of health data from various medical devices to a computerized system, connecting healthcare professionals with patients. These devices allow healthcare providers to monitor patients' health remotely (for outpatients) or in the hospital (for inpatients) and streamline workflows, thereby improving patient health outcomes \cite{huang2023internet,hameed2021systematic,al2020intelligence}.

The IoMT and the Internet of Things (IoT) devices share similar technological protocols, including wireless communication protocols such as MQTT (Message Queue Telemetry Transport), CoAP (Constrained Application Protocol), Bluetooth Low Energy (BLE), Zigbee, Wi-Fi, and TCP/IP \cite{de2019iot}. Unlike general IoT devices, IoMT systems are designed with machine intelligence, microcontrollers, and sensors to reduce the need for human intervention in standard monitoring and healthcare procedures \cite{bajaj2021healthcare}, \cite{khatkar2020overview}, \cite{saxena2022internet}. 

\begin{figure}[htbp]
\centerline{\includegraphics[width=8cm,height=5.3cm]{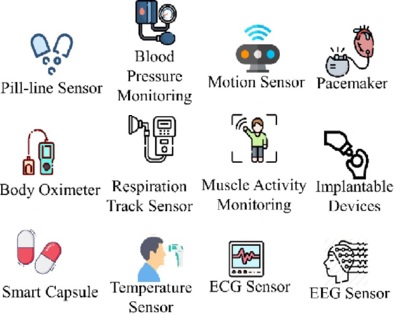}}
\caption{Examples of IoMT devices.}
\label{devices}
\end{figure}

The ecosystem of IoMT comprises devices and software applications that facilitate convenient healthcare delivery, transmitting data through a network to a cloud server where it is processed and used to trigger actions or provide insights through a user interface \cite{khatkar2020overview}. Some innovative developments in IoMT range from devices as simple as thermometers, inhalers, and insulin pumps to biometric rings and advanced robotic systems~\cite{dzamesi2025review,koutras2020security}. Figure~\ref{devices} shows an example of some IoMT devices.

From a business perspective, the IoMT is not trending; however, it exhibits a growth phenomenon. The market projection shows a significant growth in the IoMT globally, and is expected to reach hundreds of billions of dollars in the next few years. According to the HealthTech Research Group, the  IoMT will reach \$861.3 billion by 2030, indicating a 28\%  compound annual growth rate (CAGR)~\cite{grandview2023iot}. In addition, the World Health Organization (WHO) anticipates a shortage of healthcare professionals by 2030 \cite{who_healthworkforce}. This projection suggests that there will not be enough healthcare workers globally to meet the demand for medical services, thereby limiting worldwide access to quality healthcare for people. Rahan et al.~\cite{rasool2022security} also indicate that the expected increase in human life expectancy will likely boost the global population by 31\% and reach 9.8 billion by 2050. Given these trends, there is a need to develop affordable and secure healthcare solutions that address people's health demands while accommodating the dual pressures of globalization and digitization in the Internet of Medical Things (IoMT).

\begin{figure*}[htbp]
\centerline{\includegraphics[width=10.5cm,height=6.5cm]{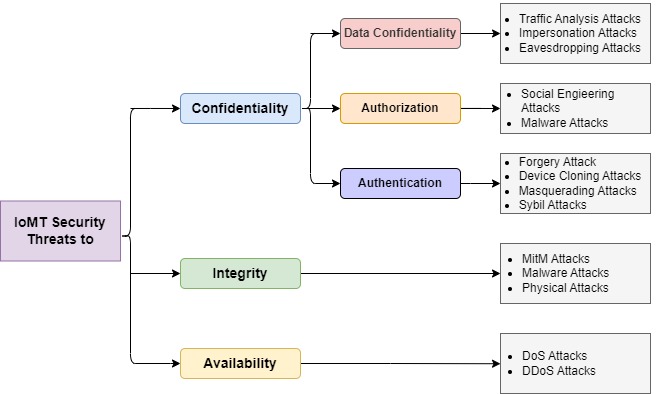}}
\caption{Security threats in IoMT network categorized by the threat to each of the confidentiality, integrity, and availability (CIA) triad.}
\label{threats}
\end{figure*}

The IoMT security landscape is characterized by various vulnerabilities that span different architectural layers and device types. Some of these vulnerabilities include poor authentication processes, such as those relying on weak or default passwords, as well as insufficient authorization controls that grant access to sensitive data and accounts. The lack of proper encryption for data in transit and data at rest presents a critical issue that exposes sensitive PHI to eavesdropping and interception \cite{vakulov2023protecting}. Many devices also suffer from insecure firmware, irregular update mechanisms, and outdated software components with known vulnerabilities. Physical security is also a vulnerability, with some devices not hardened against tampering \cite{lopez2023comprehensive}. Communication protocols used in IoMT, like Bluetooth Low Energy (BLE), Zigbee, MQTT, and CoAP, are just as vulnerable to attack if not implemented securely. The vulnerabilities attract a variety of attacks, such as malware infections (like ransomware), Distributed Denial of Service (DDoS) attacks that compromise service availability, and Man-in-the-Middle (MitM) attacks that eavesdrop on communications~\cite{al2020intelligence,elsayed2025cryptodna}. Figure~\ref{threats} shows the major IoMT security threats~\cite{papaioannou2022survey}.

From the countermeasures and vulnerabilities adopted by most studies to develop working principles for enterprise IT environments often prove inadequate, these unique challenges of IoMT:
\begin{itemize}
\item \textit{Resource Constraints:} The majority of IoMT devices, especially wearables and implants, have limited processing power, memory, and battery life, making it impossible to run complex cryptographic algorithms or heavyweight intrusion detection software designed for high-performance servers or desktops \cite{binbusayyis2022investigation}. This intrinsic trade-off between the need for strong security and hardware limitations affects the design of  security mechanisms.
    \item \textit{Heterogeneity and Lack of Standardization:} The enormous diversity of devices, manufacturers, communication protocols, and data transfer within the IoMT ecosystem makes it difficult to apply uniform security policies and solutions \cite{gudala2019leveraging}. These interoperability challenges further complicate security integration.
    \item \textit{Scalability:} Many connected IoMT devices pose severe scalability challenges for centralized security management and monitoring systems \cite{chatterjee2023approach}.
    \item \textit{Legacy Systems:} Health care environments often contain legacy devices not designed for network connectivity or security, but are being networked, bringing in severe vulnerabilities \cite{hernandez2023artificial}. These devices can be difficult or impossible to patch or update.
    \item \textit{Real-time Requirements:} Many IoMT applications, such as critical patient monitoring devices or remote control of therapeutic devices, have rigorous real-time performance expectations \cite{liaqat2020sdn}.
    \item \textit{Dynamic Environments:} The IoMT environment is typically dynamic, as devices frequently enter and exit the network, and user contexts, such as location, change often \cite{deng2019overview}. Therefore, static security policies are inadequate.
\end{itemize}

Among these vulnerabilities and attack vectors statistics provided by the 2024 Elastic Global Threat Report~\cite{elastic2024threat}, published by Cybersecurity Ventures, indicate that the leading threat concerns targeting Internet of Medical Things (IoMT) systems has been determined to be Distributed Denial of Service (DDoS) attacks and malware attacks represent the most significant cybersecurity threats confronting the IoMT ecosystem, among other potential threats. These attacks pose a serious risk to the integrity, availability, and confidentiality of IoMT systems~\cite{al2023cybersecurity},~\cite{elhoseny2021security},~\cite{hireche2022security}. 

Due to the critical aspect of IoMT devices' availability, especially for life support IoMT devices, this paper targets the detection of DDoS attacks on IoMT networks to protect and save lives from critical threats. 

The resource limitations of IoMT networks are a significant reason for the successful launches of DDoS attacks on IoMT devices; they also contribute to security limitations due to the inability to employ security software, which increases the risk of IoMT devices being compromised. Thus, in this paper, we aim to employ a budget-friendly machine learning model based on the extreme learning machine (ELM). This robust and powerful approach requires minimal resources and can achieve a significantly high accuracy in detecting DoS and DDoS attacks. Thus, the main contribution of this paper is to empirically investigate the capability of employing the ELM to detect different DDoS attacks on IoMT network devices, focusing on accuracy and low-budget implementation, which is compatible with the limited resources of the IoMT network.

\begin{figure}[htbp]
\centerline{\includegraphics[width=8.5cm,height=6cm]{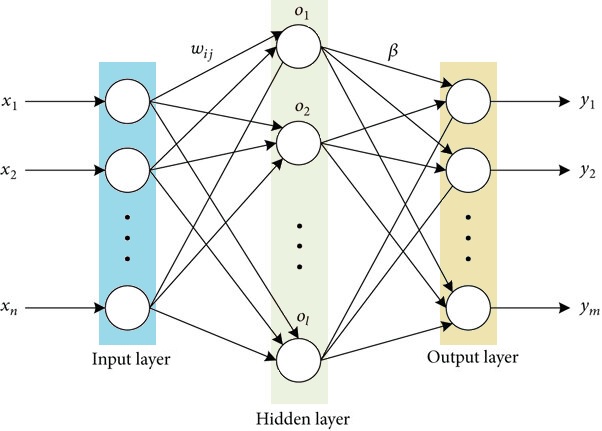}}
\caption{The ELM topology.}
\label{elm_topology}
\end{figure}

\begin{table*}[htbp]
\caption{Comparison Between ELM and the Other Artificial Neural Networks}
\begin{center}
\begin{tabular}{|p{2.5cm}|p{3.1cm}|p{3.1cm}|p{3.1cm}|p{3.1cm}|}
\hline
\textbf{Aspect} & \textbf{ELM}& \textbf{FNN}& \textbf{BPNN} &\textbf{1D-CNN}\\
\hline
Architecture&	Single hidden layer feedforward network&	Multi layer fully connected network&	Multi layer with backpropagation&	1D filters: Convolutional, pooling, and dense layers\\
\hline
Weight Initialization&Input weights and biases randomly assigned&	Initialized and updated during training	&Initialized and updated through backpropagation	&Initialized and updated via backpropagation\\
\hline
Training Algorithm&	Analytical solution using Moore–Penrose pseudoinverse&	Gradient-based &Backpropagation with gradient descent	&Backpropagation with gradient descent\\
\hline
Training Speed&	Very fast (non-iterative)&	Moderate&	Slower (iterative, often many epochs)&	Slower (requires convolution and tuning)\\
\hline
Learning Type&	Batch learning only	&Batch or online &	Batch or online&	Batch or mini-batch learning\\
\hline
Generalization Ability	&Good with proper parameters&	Depends on training and architecture & Requires regularization to prevent overfitting &Strong with large data, spatial pattern learning\\
\hline
Interpretability&	Moderate	&Moderate	&Low	&Low\\
\hline
Data Type& Tabular, structured data	&Tabular, structured data	&Tabular, structured data	&Sequential, time series data\\
\hline
Use Cases&	Fast classification, intrusion detection, anomaly detection	&General classification/regression tasks&	General ML tasks, especially small/medium datasets	&Time-series classification, signal analysis\\
\hline
Parameter Tuning&	Fewer parameters (number of neurons)	&Requires tuning of layers, units, activation	&Requires tuning of layers, learning rate	&Requires tuning of kernel size, stride, filters\\
\hline
Scalability&	Limited scalability for very large data	&Scalable with regularization and optimization	&Scalable with GPU, but slow	&Highly scalable with GPU support\\
\hline
Usage for IoMT Edge&	Highly suitable: fast, lightweight, low compute	&Moderately suitable: need optimization&	Less suitable: slow training, higher overhead	&Suitable for time series sensor data, more resource heavy\\
\hline

\end{tabular}
\label{compare_ELM}
\end{center}
\end{table*}

\section{Extreme Learning Machine}
The extreme learning machine (ELM) was proposed in 2006 by Huang et al.~\cite{huang2006extreme} as a simple learning algorithm for the Single Hidden Layer feedforward neural network (SLFN) with N hidden nodes~\cite{matias2014learning,hornik1991approximation,nayak2022extreme} where the SLFN is a simplest type of feedforward network which architecture design consists of a single-hidden-layer perceptron~\cite{maan2016survey}. 

The ELM topology is shown at Figure~\ref{elm_topology} where $X = {[x_{1}, x_{2},...., x_{n}]}^T$ is the input vector, $\mathit{w_j}$ is the weights from inputs to hidden node $\mathit{j}$, $n$ is the number of features, and .$b_{j}$ is the bias of the hidden node $j$. For each hidden node $j = {1, 2, ...., L}$, the output of the hidden neuron $o_{j}$ is calculated by:

\begin{equation}
    o_{j} = g(\textbf{w}_{j} . \textbf{X} + b_j)
\end{equation}
\noindent
where $g(.)$ is the activation function. For all the training samples, the output can be calculated by:
\begin{equation}
   \textbf{Y} = \textbf{H} \beta
\end{equation}
\noindent where H is the hidden layer output matrix for all the training samples where:
\begin{equation}
   \textbf{H} = g(\textbf{XW} +\textbf{b})
\end{equation}
\noindent
where $\textbf{X} \in \mathbb{R}^{N \times n}$, $\textbf{W} \in \mathbb{R}^{n \times L}$,  $\textbf{b} \in \mathbb{R}^{1 \times L}$, $\textbf{H} \in \mathbb{R}^{N \times L}$. $N$ is the number of samples, and $L$ is total number of hidden nodes, $\beta$ is the weights from the hidden to the output layer, where:
\begin{equation}
    \beta = \textbf{H}^{T} T
\end{equation}
\noindent
$\textbf{T} \in \mathbb{R}^{N \times m}$ is the target (label) matrix, and $\textbf{H}^{T}$ is the Moore–Penrose pseudoinverse of $\textbf{H}$~\cite{barata2012moore}.

Unlike traditional neural networks, which employ iterative gradient-based methods (such as backpropagation) to adjust all network weights, the ELM assigns random weights and biases to its hidden layer. It keeps them constant, learning only the output weights using a closed-form solution (usually least squares via the pseudoinverse). Such network design enables ELMs to train much faster compared to traditional neural networks, often with comparable generalization performance across numerous problems~\cite{reddy2023mdc}. Table~\ref{compare_ELM} shows a comparison between the ELM, the feedforward neural network (FNN), the back propagation neural network (BPNN), and the 1D convolutional neural network (1D-CNN)~\cite{huang2006extreme,rumelhart1986learning,kiranyaz2015real,elsayed2018deep,al2012privacy}.

The ELM has shown a significant performance in different applications and tasks, including healthcare applications, including diabetes prediction~\cite{elsayed2022early}, electrocardiogram (ECK/EKG) classification~\cite{elsayed2020simple}, and electroencephalogram (EEG) classification~\cite{ding2015deep}. Also, it shows significant success at sparse representations for images~\cite{cao2016extreme}, action recognition~\cite{kuang2017extreme}, image classification~\cite{cao2013image}, and intrusion detection in networks~\cite{cheng2012extreme,singh2015intrusion}.

\begin{figure}[t]
\centerline{\includegraphics[width=8.5cm,height=1.5cm]{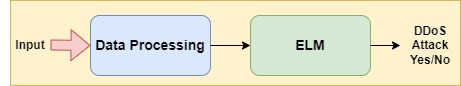}}
\caption{The proposed ELM based approach to detect DDoS attacks on IoMT devices.}
\label{proposed_model}
\end{figure}

\section{Methodology}
In this paper, we employed the ELM to detect DDoS attacks on the IoMT network devices. The proposed approach aims to achieve high detection accuracy while maintaining a small implementation budget and accelerating the detection speed. The proposed model is shown in Figure~\ref{proposed_model}. The proposed approach consists of three main stages: data input, data processing, and classification using the ELM. The proposed methodology is designed to provide an accurate detection of malicious network traffic patterns while maintaining rapid processing. The proposed methodology compromises the availability and reliability of IoMT systems.

The first stage involves collecting input data, which includes real-time or recorded network traffic from IoMT devices. The data consists of a wide range of features that describe the characteristics of the network flows, including the packet size, protocol type, flow duration, source and destination IP addresses, port numbers, and the number of connection requests. These features form the basis for identifying potential DDoS attack patterns embedded within the network traffic.

After collecting the data, it goes through a processing stage. The data processing stage involves cleaning the data to remove any missing, duplicate, or noisy data samples that may affect the performance and the learning process of the ELM detection model. Feature extraction and selection techniques are then applied to identify the most relevant attributes for detecting DDoS attacks. Then, the selected features are normalized to enhance the learning efficiency and stability of the ELM model. Finally, the data is labeled for the normal and DDoS attacks. 

Then, the learning stage starts by utilizing the ELM as a classifier. In this approach, the input weights and biases are randomly initialized, and the hidden layer outputs are computed using an activation function. The final stage involves generating the output of the classification model, which provides a binary decision indicating whether a DDoS attack is detected. 

\section{Experiments and Results}

\subsection{Dataset}
In this paper, we employed the CICIoMT2024 dataset benchmark~\cite{dadkhah2024ciciomt2024}. We selected the CICIoMT2024 dataset benchmark for our experiments due to several factors, including that the CICIoMT2024 dataset is a realistic benchmark dataset that enables the development and evaluation of solutions that focus specifically on IoMT security. Additionally, this dataset comprises 18 distinct DDoS attacks against an IoMT testbed comprising 40 IoMT devices, with 25 real devices and 15 simulated devices. The 18 types of DDoS attacks are categorized into the main five categories: Distributed Denial of Service (DDoS), Denial of Service (DoS), Reconnaissance (Recon), Message Queuing Telemetry Transport (MQTT) attacks, and Spoofing. Table~\ref{attacks} shows the main attacks and the sub-attacks for each of the five main categories. The dataset consists of 476,150 data samples.

\begin{table}[htbp]
\centering
\caption{The CICIoMT2024 dataset benchmark attack categories and their sub-categories}
\renewcommand{\arraystretch}{1.2}  
\begin{tabular}{|>{\raggedright\arraybackslash}p{3cm}|>{\raggedright\arraybackslash}p{3cm}|}
\hline
\textbf{Category} & \textbf{Sub-category} \\
\hline
DDoS & SYN Flood \newline TCP Flood \newline ICMP Flood \newline UDP Flood \\
\hline
DoS & SYN Flood \newline TCP Flood \newline ICMP Flood \newline UDP Flood \\
\hline
Recon & Ping Sweep \newline Vulnerability Scan \newline OS Scan \newline Port Scan \\
\hline
Spoofing & ARP Spoofing \\
\hline
MQTT & Malformed Data \newline DoS Connect Flood \newline DDoS Connect Flood \newline DoS Publish Flood \newline DDoS Publish Flood \\
\hline
\end{tabular}

\label{attacks}
\end{table}

\begin{table}[htbp]
\caption{The Proposed ELM Based Approach Testing Results}
\begin{center}
\begin{tabular}{|c|c|}
\hline
\textbf{Metrics} & \textbf{Result}\\
\hline
Accuracy (\%)&94.87\% \\
Precision &0.95\\
Recall &0.95\\
F1-Score&0.95\\
AUC-ROC&0.945\\
\hline
\end{tabular}
\label{tab1}
\end{center}
\end{table}

\begin{figure}[t]
    \centering
    \includegraphics[width=6cm,height=5cm]{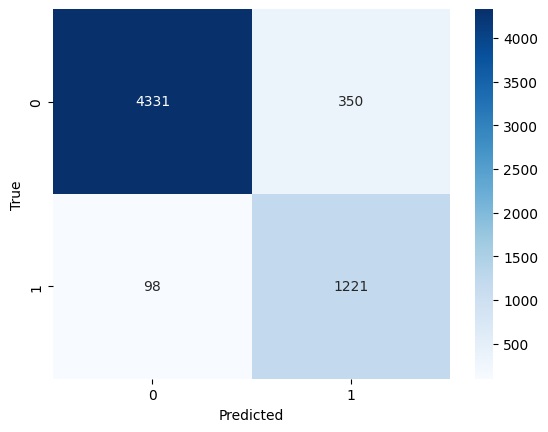}
    \caption{The confusion matrix of the ELM model for detecting DDoS attacks on the CICIoMT2024 dataset benchmark.}
    \label{confusion_matrix}
\end{figure}

\subsection{Data Processing}
For data processing, we performed data cleaning, removing missing values. In addition, we used it for testing and training. Non-benign traffic was labeled as 1 (indicating a DDoS attack), while benign traffic was marked 0.

To reduce dimensionality and enhance model effectiveness, the correlation of each feature with the target class was determined. Features with an absolute correlation value of less than 0.02 were discarded. Only the most predictive features related to DoS and DDoS detection remained.

\begin{table*}[htbp]
\centering
\caption{An Analytical Comparison with the State-of-the-Art Modeld for DDoS Detection}
\begin{tabular}{|p{2.0cm}|p{2.0cm}|p{2.0cm}|>{\centering\arraybackslash}p{1.2cm}|>{\centering\arraybackslash}p{1.2cm}|>{\centering\arraybackslash}p{1.2cm}|>{\centering\arraybackslash}p{1.2cm}|p{2.cm}|} \hline 

\textbf{Study} & \textbf{Model Type} & \textbf{Dataset} & \textbf{Accuracy (\%)} & \textbf{Precision} & \textbf{Recall} & \textbf{F1-Score} & \textbf{Specific Findings} \\ \hline 

\textbf{Our Work} & ELM & CICIoMT24  & 95.0 & 0.94 & 0.95 & 0.95 & Baseline model for comparison \\ \hline 

Neto et al.~\cite{tacsci2024deep}&CICIoV2024 &Logistic Regression&95\%& 0.74&0.68&0.63&Logistic Regression showing weak detection despite high accuracy\\
\hline

Kharoubi et al.~\cite{kharoubi2025network} & Convolutional Neural Network (CNN) & CICIoT2023 (Medical subset) & 99.1 & 0.986 & 0.99 & 0.99 & Focused on IoT security using CNN-based NIDS \\ 
\hline 

Canavese et al.~\cite{canavese2024security}&Random Forest&	CICIoT2023 &95.7\% &0.28 &	0.65&0.33&	High accuracy but precision and recall very low\\
\hline
Ansar et al.~\cite{ansar2024cutting} & Hybrid CNN & CICIoMT24 + IoMT-TON & 97.8 & 0.972 & 0.972 & 0.968 & For IoT security application \\ \hline 

Chandekar et al.~\cite{chandekar2025enhanced} & Ensemble AI & CICIoMT2024 & 96.8 & 0.965 & 0.971 & 0.968 & Enhanced anomaly detection in IoMT networks \\ \hline 

Gueriani et al.~\cite{gueriani2024enhancing} & Hybrid CNN-LSTM & CICIoT2023, CICIoMT24 & 99.3 & 0.98 & 0.991 & 0.99 & Enhanced IoT security with CNN-LSTM-based IDS \\ \hline 

Fernández Maimó et al.~\cite{urooj2021ransomware} & Dynamic analysis system & CICIoMT24 & 97.2 & 0.968 & 0.975 & 0.971 & Focused on ransomware detection in clinical environments \\ \hline

\end{tabular}
\label{models_compare}
\end{table*}

The dataset was split into 80\% training and 20\% test sets \( X_{\text{train}}, Y_{\text{train}}\). The features were normalized using z-score normalization via StandardScaler so that all input features have the same scaling~\cite{fei2021z}.

\subsection{Model Implementation and Evaluation}

The ELM classifier was implemented based on the model architecture components:
\begin{itemize}
    \item One hidden layer with a tunable number of neurons.
    \item Random initialization of input weights and bias.
    \item Activation functions: tanh, sigmoid, and radial basis function (RBF)~\cite{elsayed2018empirical,buhmann2000radial}.
    \item Output weights were computed through the Moore-Penrose pseudoinverse to obtain a closed-form solution.
\end{itemize}

The model was encapsulated in a special ELMWrapper class to facilitate integration with GridSearchCV. The hyperparameters utilized the number of neurons and the activation function type to optimize a 5-fold cross-validation approach.

We assessed our model's performance through binary classification. This involved training it on the datasets to differentiate between normal and attack traffic. We used a series of assessment metrics to evaluate its effectiveness. 

The model recorded an overall accuracy of 94.87\%, reflecting its high potential to distinguish between malicious and normal traffic in IoMT networks. The classification report shows a high accuracy and recall rate of 0.95 for the two classes (benign and attack), indicating that the model effectively and efficiently minimizes false negatives and false positives, as shown in the confusion matrix~\ref{confusion_matrix}. Table~\ref{results1} shows the ELM testing results. The ELM model performed exceptionally well while using a small implementation budget. It also recorded a precision of 0.95, a recall of 0.95, and an F1-score of 0.95. Our proposed approach proved to outperform most traditional machine learning models. In addition to achieving high accuracy, the ELM model demonstrated high training speeds and quick time complexity, making it suitable for real-time applications. This feature benefits IoMT systems requiring real-time threat detection with limited resources. The results further confirm the effectiveness of the ELM-based detection system for classifying malware and DDoS attacks.

A comparative analysis of various machine learning algorithms was conducted to assess the strengths and weaknesses of the ELM model in relation to traditional methods. The performance of our ELM model was compared with that of other methods, with a particular focus on models tested on the CICIoMT24 database for fair comparison. The study identifies the critical dimensions for IoMT security solutions, including detection accuracy, computational efficiency, precision, and recall. Table~\ref{models_compare} shows a comparison between the proposed ELM-based approach and other state-of-the-art models. The CNN-LSTM hybrids and deep neural networks used in different studies have high-accuracy models; however, our ELM model classifier is more applicable to IoMT systems. Also, deep learning models, such as those developed by Ansar et al.~\cite{ansar2024cutting} and Gueriani et al.~\cite{gueriani2024enhancing}, have marginally higher accuracy rates (97–98\%). Still, they are more resource-intensive, they require more advanced training and processing power, which are not feasible for medical devices with limited resources. For the more suitable models for IoMT devices from the implementation resources aspect, Neto et al.~\cite{tacsci2024deep} and Canavese et al.~\cite{canavese2024security} models showed high accuracy; however, they have a significantly low precision and recall, evidencing poor per-class performance. Thus, the ELM model was distinguished by:

\begin{itemize}
    \item Single-pass learning speeds for training.
    \item Its lower memory usage makes it simple to install on IoMT devices.
    \item Tuning is facilitated due to the reduced number of training epochs and hyperparameters.
    \item Real-time intrusion detection offers a high rate of accuracy, 95\%, which is crucial in healthcare facilities.
    \item Maintaining high precision, recall, and F1-Score.
\end{itemize}

Our ELM model offers better training efficiency because of its single-pass learning mechanism, unlike the iterative backpropagation used by models like CNNs and LSTMs, which significantly increases training time and computational load. Furthermore, the ELM model also provides low CPU and memory usage, hence it is highly suitable for deployment on resource-limited edge devices. 

\section{Conclusion}
This paper proposes an  Extreme Learning Machine (ELM) approach for detecting DDOS attacks on Internet of Medical Things (IoMT) systems within connected medical networks. The proposed ELM model was applied and tested using real-world simulated attack traffic from the CICIoMT24 dataset. The proposed ELM approach achieved significant classification performance in detecting both benign and actual attacks with 95\% accuracy, requiring minimal training time and implementation resources. The ELM model makes it a suitable option for real-time attack detection in resource-constrained healthcare systems.

\bibliographystyle{ieeetr}
\bibliography{references_ELM}

\end{document}